\documentclass[prb,superscriptaddress,twocolumn]{revtex4-2}
\usepackage{graphicx}
\usepackage{times,amsmath}
\usepackage{epsfig}
\usepackage{color}
\usepackage{graphicx}
\usepackage{dcolumn}
\usepackage{bm}
\usepackage{bookmark}
\usepackage{tabularx}
\usepackage{hyperref}
\usepackage{multirow}
\usepackage{array,mathtools,amssymb,booktabs,makecell}
\hypersetup{colorlinks=true, citecolor=blue, filecolor=blue, linkcolor=blue, urlcolor=blue}

\usepackage{float}
\usepackage{adjustbox}
\usepackage{upgreek}
\usepackage{soul,xcolor}
\setstcolor{red}

\makeatletter
\let\saved@includegraphics\includegraphics
\AtBeginDocument{\let\includegraphics\saved@includegraphics}
\renewenvironment*{figure}{\@float{figure}}{\end@float}
\makeatother

\begin{document}

\title{Coexistence of 3D and quasi-2D Fermi surfaces driven by orbital selective Kondo scattering in UTe$_2$}

\author{Byungkyun Kang}
\email[]{bkang@udel.edu}
\affiliation{College of Arts and Sciences, University of Delaware, Newark, DE 19716, USA}




\author{Myoung-Hwan Kim}
\affiliation{Department of Physics and Astronomy, Texas Tech University, Lubbock, Texas 79409, USA}

\author{Chul Hong Park}
\affiliation{Research Center of Dielectric and Advanced Matter Physics, Pusan National University, Busan 46240, Republic of Korea}

\begin{abstract}
The 3D Fermi surface, along with a chiral in-gap state and a Majorana zero energy state, is suggested to play a crucial role in the topologically nontrivial superconductivity in UTe$_2$.
However, conflicting experimental observations of the 2D Fermi surface raise questions about topological superconductivity.
By combining ab initio many-body perturbation GW theory and dynamical mean-field theory based on Feynman diagrams, we discovered the coexistence of two orbital dependent Fermi surfaces, both centered at the $\Gamma$ point in the Brillouin zone, which are heavily influenced by the orbital-selective Kondo effect.
At high temperature, both Fermi surfaces exhibit 3D nature with weak spectral weight due to incoherent Kondo hybridization.
Upon cooling down to 25 K, due to the pronounced Kondo coherence,
while one Fermi surface remains a robust 3D Fermi surface,
the other transforms surprisingly into a quasi-2D Fermi surface, which should be responsible for the experimental observation of 2D character.
Our results suggest that the 3D Fermi surface should exist at low temperature for the topological superconductivity. Our findings call for further investigation of the interplay between the two orbital-dependent $\Gamma$-centered Fermi surfaces.
\end{abstract}


\maketitle

\section*{Introduction} The plausibility of topological superconductivity in UTe$_2$ has been a topic of considerable contention, contingent upon the presence of a three-dimensional (3D) Fermi surface. Recent scanning tunneling microscopy/spectroscopy measurements have revealed the presence of a chiral in-gap state and Majorana zero energy states in UTe$_2$~\cite{Jiao2020}, providing evidence for the existence of a 3D Fermi surface.
Two studies using angle-resolved photoemission spectroscopy, while presenting contrasting results, have suggested the existence of a possible 3D Fermi surface measured at 20 K~\cite{Fujimori2019,lin_prl2020}.
This is further supported by quantum oscillation measurements below 1.2 K, which show coexistence of 3D and 2D Fermi surface pockets~\cite{christopher_prl2023}. However, recently, conflicting results have been reported. Fermi surface measurements using the de Haas-van Alphen effect below 200 mK~\cite{eaton_ncom2024} and high magnetic field magneto-conductance oscillations below 4.2 K~\cite{weinberger_prl2024} identified only quasi-two-dimensional (quasi-2D) Fermi surface. 
One of main ingredient for the topological superconductivity is 3D Fermi surface since quasi-2D cylindrical Fermi surface can't support topologically nontrivial state.
These complicated phenomena are not yet explained by theoretical studies. 
Tight binding model~\cite{jun_prb2021}, density functional theory (DFT)~\cite{Fujimori2019}, and DFT combined with dynamical mean field theory (DMFT)~\cite{hong_arxiv2023} all predicted a 3D Fermi surface centered at the $\Gamma$ point in the Brillouin zone. On the other hand, DFT+$U$ calculations~\cite{jun_prl2019,lin_prl2020,yuanji_prl2019} have reported only non-$\Gamma$-centered quasi-2D cylindrical sheet of the Fermi surface. 
There is a pressing demand for a microscopic careful examination of Fermi surface topology from a first-principles many-body perspective to understand topological superconductivity in UTe$_2$. In this work, it is shown that orbital dependent Kondo hybridization plays an important role in the complex reconstruction of both $\Gamma$-centered 3D and quasi-2D Fermi surfaces, on top of a non-$\Gamma$-centered quasi-2D cylindrical sheet of the Fermi surface.

The enigmatic properties of UTe$_2$ extend beyond its unconventional superconductivity, as evidenced by its re-entrant behavior that is contingent upon the orientation of the applied magnetic field \cite{Ran2019,Ran2020,Aoki2019review}. This intriguing phenomenon can be elucidated by the anisotropic electronic structure, as demonstrated by the orbital selective Kondo effect \cite{byung_ute2,eo_prb2022}. Furthermore, under high pressure, UTe$_2$ has been observed to exhibit long-range magnetic ordering \cite{thomas_scadv2021,dexin_jpsj2021,braithwaite_comphys2019,sheng_prb2020}. This has been attributed to the momentum-dependent $f$-$d$ Kondo hybridization. The coherent Kondo hybridization under high pressure leads to the delocalization of Bloch-like U-5$f$ quasiparticles and the van Hove singularity, which is the origin of the magnetic moment \cite{kang_ukondo}. These studies demonstrate that the Kondo effect has a significant impact on the electronic structure of UTe$_2$.

\begin{figure*}[ht]
\centering
\includegraphics[width=0.99
\textwidth]{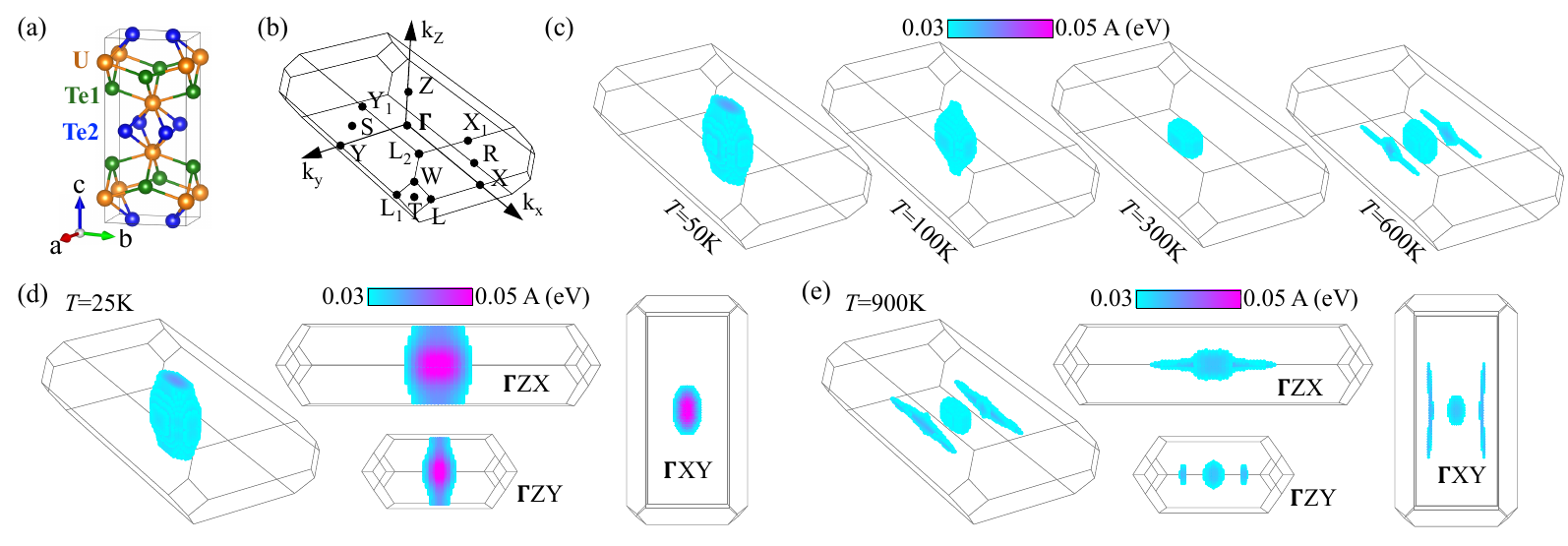}
\caption{\label{Fig_2DFS} (a) Crystal structure of UTe$_2$. (b) The first Brillouin zone. (c) Te2-5$p$ state of $|j= 3/2,j_{z}= -3/2\rangle$ projected Fermi surface in the Brillouin Zone.
Te2-5$p$ state of $|j= 3/2,j_{z}= -3/2\rangle$ projected Fermi surface and its cross section at $T$=25 K (a) and $T$=900 K (b). In (c), (d), and (e), the same spectral weight, as noted by the color bar in (d), was used.
}
\end{figure*}

In the Kondo lattice UTe$_2$, the intense local magnetic moment of the U-5$f$ orbitals is influenced by the Kondo hybridization with conduction electrons at the Fermi level. Consequently, the shape and properties of the Fermi surface are significantly influenced by the Kondo scale and the specific orbitals involved.
To describe the Kondo effect, a phenomenon arising from strong electron correlation, an ab-initio many-body method is necessary for accurately depicting the Fermi surface.
The ab-initio many-body approach has yielded a calculated static $U(\omega=0)$ value of 3.29 eV for UTe$_2$~\cite{byung_ute2,kang_ukondo}, indicating a moderate strength of Mottness. Moreover, it has been shown that UTe$_2$ does not display a pseudogap up to 6000 K~\cite{byung_ute2}, suggesting its similarity to a Hund's metal, where the exchange interaction $J$ dominates over $U$~\cite{hazra_prl2023,byung_lanio2}.
Thus, the presence of metallic U-5$f$ spectral weight in the vicinity of the Fermi level remains prominent at elevated temperatures, resulting in incoherent Kondo scattering. 
In general, Kondo scattering becomes more coherent as the temperature decreases. However, the evolution of the Fermi surface resulting from orbital selective Kondo scattering is unknown. 

A common characteristic shared by unconventional superconductors based on copper and iron is the quasi-2D Fermi surface, which strongly restricts the possible pairing symmetries \cite{bollinger_sst2016,qimiao_natrev2016}. In the case of UTe$_2$, the quasi-2D Fermi surface is likely to play a significant role in the unconventional superconductivity. Although the quasi-2D Fermi surface does not contribute to the nontrivial topological invariant, the emergence of a 3D Fermi surface could potentially host topological superconductivity through odd-parity pairings~\cite{hong_arxiv2023}.

\section*{Results}
In this work, we reveal, through ab-initio many-body calculations that explicitly include the Kondo effect, the coexistence of 3D and quasi-2D Fermi surfaces centered at the $\Gamma$ point at low temperatures, due to the orbital selective Kondo effect. 
In the direction of the $c$ axis, incoherent Kondo hybridization primarily involving Te-5$p$ and U-5$f$ orbitals gives rise to a 3D Fermi surface centered at the $\Gamma$ point at high temperatures. Interestingly, as the temperature decreases, the quasi-2D Fermi surface is found to emerge due to the enhancement of the coherent Kondo hybridization.
Additionally, there is an incoherent dispersive 3D Fermi surface comprising U-6$d$ states at high temperature. Upon cooling, this 3D Fermi surface becomes more coherent with pronounced spectral weight.
Our results support the quantum oscillation measurements that demonstrate the coexistence of 2D and 3D Fermi surfaces~\cite{christopher_prl2023}.
We examined the electronic structure of UTe$_2$ by using ab-initio linearized quasiparticle self-consistent GW + dynamical mean field theory (LQSGW+DMFT) method~\cite{choi2019comdmft}.
This method enables us to accurately determine the temperature-dependent electronic structure of strongly correlated materials without any subjective judgment. 
Aside from adopting an experimental lattice parameter, we explicitly compute all other quantities, such as double-counting energy and Coulomb interaction tensor.
The LQSGW+DMFT method has been used to investigate the Kondo effect in compounds based on actinides and lanthanides~\cite{byung_usbte,kang_ukondo,byung_ute2,kang_ndnio2}.

\begin{figure*}[ht]
\centering
\includegraphics[width=0.99
\textwidth]{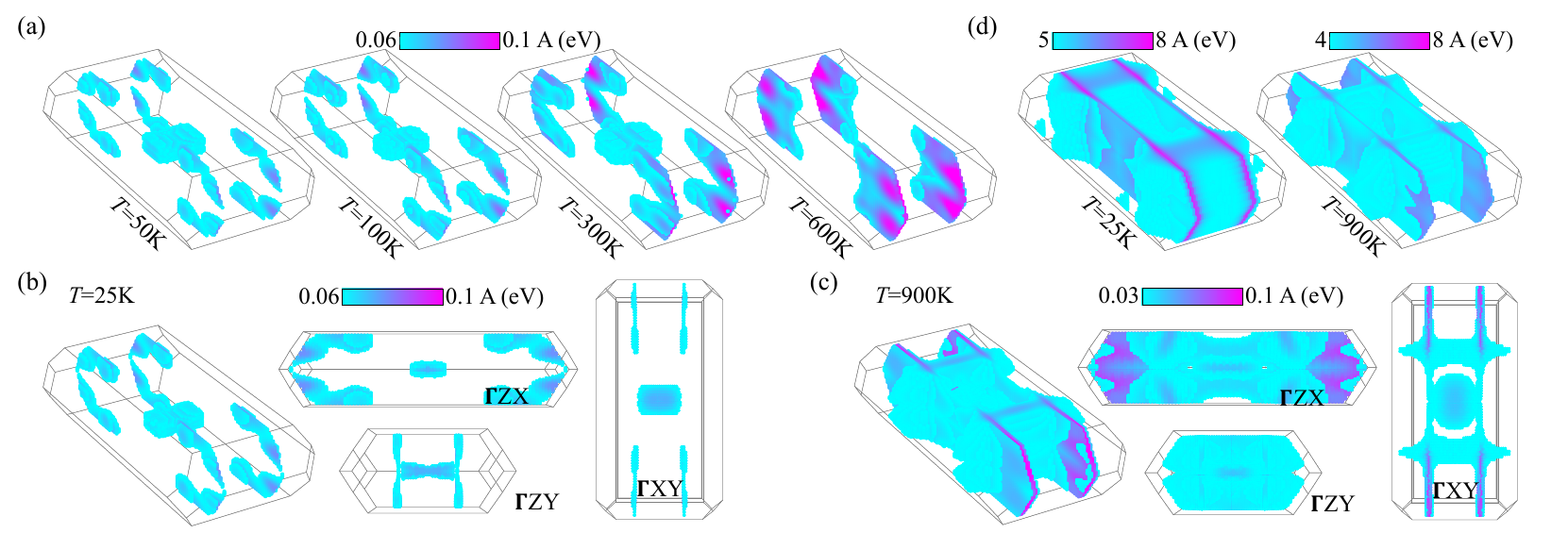}
\caption{\label{Fig_3DFS} (a) U-6$d$ states of $|j= 3/2,j_{z}= -3/2\rangle$ projected Fermi surface in the Brillouin Zone.
U-6$d$ states of $|j= 3/2,j_{z}= -3/2\rangle$ projected Fermi surface and its cross section at $T$=25 K (b) and $T$=900 K (c). (d) The Fermi surface of UTe$_2$ at $T$=25 K and 900 K.
}
\end{figure*}

\subsection*{Transition from 3D to quasi-2D Fermi surface of Te2-5$p$} The crystal structure of the orthorhombic phase (Immm) UTe$_{2}$~\cite{ikeda2006single} is described in Fig.~\ref{Fig_2DFS}a. The high symmetry k-points in the first Brillouin zone are shown in Fig.~\ref{Fig_2DFS}b.
UTe$_2$ consists of two Te atoms with Wyckoff positions 4j (Te1) and 4h (Te2). Te2-5$p$-orbitals were shown to be responsible for the Kondo effect along the $c$ axis~\cite{byung_ute2}.
Here, we show that the Fermi surface projected by the Te2-5$p$ state of the $|j= 3/2,j_{z}= -3/2\rangle$ in the Brillouin zone significantly changes with a decrease of temperature, as shown in Fig.~\ref{Fig_2DFS}c. It is remarkable that as the temperature is lowered, the Fermi surface transforms from the closed 3D one to the open 2D one. At high $T$ of 600 K, a 3D oblate Fermi surface forms at $\Gamma$, whereas
at low $T$ of 50 K, it transforms to be quasi-2D cylindrical.
In Fig.~\ref{Fig_2DFS}d and e, more details are compared by the cross sections of the Fermi surface between $T$ = 900 K and 25 K. Both Fermi surfaces exhibit densely packed spectral weight, unlike 3D pocket Fermi surface observed in quantum oscillations measurements~\cite{christopher_prl2023}.
At 900 K, the spectral weight on the Fermi surface is broadened and incoherent, whereas at 25 K, the spectral weight is more increased along the $\Gamma$-Z high symmetry direction.
As an orbital selective Kondo lattice, the Te2-5$p$ states of $|j= 3/2,j_{z}= \pm3/2\rangle$ hybridize with the U-5$f$ states of $|j= 5/2,j_{z}= \pm5/2\rangle$, which are responsible for the Kondo effect along the $c$ axis~\cite{byung_ute2}.
On the Fermi surface project by U-5$f$ states of $|j= 5/2,j_{z}= \pm5/2\rangle$, the spectral weight is dispersive across the entire Brillouin zone. 
This indicates that the flat U-5$f$ electrons lead to the Kondo scattering of Te2-5$p$, and the 3D shape of the Fermi surface is mostly contributed by Te2-5$p$.

The Te2-5$p$-driven spectral weight on the Fermi surface increases at low temperature of 25 K, and 
the DOS of the U-5$f$ states of $|j= 5/2,j_{z}= \pm5/2\rangle$ at the Fermi level exhibit a progression in the formation of a quasiparticle peak at low T below 50K~\cite{byung_ute2}.
These findings suggest that the coherent Kondo resonance between the Te2-5$p$ states of $|j= 3/2,j_{z}= \pm3/2\rangle$ and the U-5$f$ states of $|j= 5/2,j_{z}= \pm5/2\rangle$ drives the transformation of the 3D Fermi surface into the quasi-2D Fermi surface.

\subsection*{Enhanced 3D Fermi surface of U-6$d$} We found that the U-6$d$ states of $|j= 3/2,j_{z}= \pm3/2\rangle$ give rise to a 3D Fermi surface centered at $\Gamma$.
Figure~\ref{Fig_3DFS}a shows that the 3D Fermi surface of the U-6$d$ states of $|j= 3/2,j_{z}= -3/2\rangle$ centered at $\Gamma$. At 900 K, the 3D Fermi surface is invisible within the given range of spectral weight. Upon cooling, the 3D Fermi surface appears due to stronger spectral weight, with a slightly varying size. 
Figure~\ref{Fig_3DFS}b shows the cross sections of the Fermi surface at 25 K. It exhibits a dumbbell shape, unlike the oval shape obtained from LDA+DMFT~\cite{hong_arxiv2023}.
As shown in Fig.~\ref{Fig_3DFS}c, the 3D Fermi surface appears at a high temperature of 900 K. The spectral weight is weak and dispersive, making it visible only at a lower range of spectral weight ($<$ 0.06 A (eV)).  

\begin{figure*}[ht]
\centering
\includegraphics[width=1.0
\textwidth]{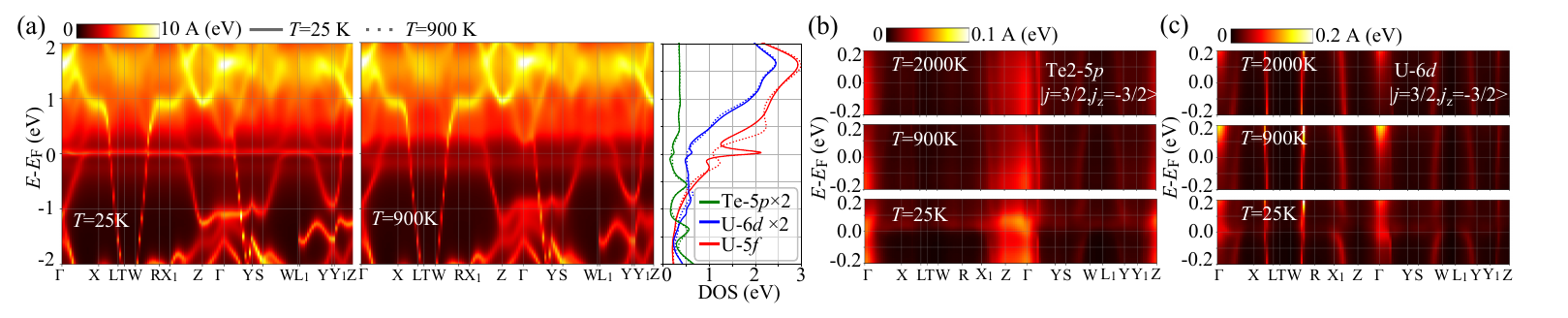}
\caption{\label{Fig_bands} (a) The calculated spectral functions and density of states for UTe$_2$ at $T$=25 K (solid lines) and 900 K (open). Te2-5$p$ state of $|j= 3/2,j_{z}= -3/2\rangle$ (b) and U-6$d$ states of $|j= 3/2,j_{z}= -3/2\rangle$ (c) projected spectral functions at $T$=2000, 900, and 25 K.
}
\end{figure*}

To determine the Fermi surface topology, we calculated the Fermi surface for all orbitals, as shown in Fig.~\ref{Fig_3DFS}d. The calculated Fermi surface is non-$\Gamma$-centered quasi-2D cylindrical sheet at $T$=900 K and consists mainly of U-6$d$ (see Fig.~\ref{Fig_3DFS}c) and dispersive U-5$f$ orbitals.
The non-$\Gamma$-centered quasi-2D cylindrical sheet of Fermi surface is similar to that obtained by the DFT+$U$ method~\cite{lin_prl2020,yuanji_prl2019}, because the majority of the cylindrical Fermi surface is composed of weakly correlated U-6$d$ and Te-5$p$ bands.
At $T$=25 K, the Fermi surface still retains its quasi-2D feature. However, the Fermi surface along the $k_x$ axis becomes more coherent, while the Fermi surface along the $k_y$ axis becomes more dispersed compared to the Fermi surface at $T$=900 K. 
The dispersion is attributed to the kink-like band structure along the $\Gamma$-X high symmetry line at $T$=25 K, which is caused by a stronger Kondo hybridization between U-6$d$ and a subgroup of U-5$f$~\cite{byung_ute2} compared to $T$=900 K, as shown in Fig.~\ref{Fig_bands}a.
The $\Gamma$ centered Te2-5$p$ projected quasi-2D Fermi surface and U-6$d$ projected 3D Fermi surface are not visible in Fig.~\ref{Fig_3DFS}d due to their relatively small spectral weight. Nonetheless, the coexistence of the 3D and quasi-2D Fermi surfaces centered at $\Gamma$ is apparent in our orbital projected Fermi surface.
 
\subsection*{Coexistence of 3D and quasi-2D Fermi surface driven by orbital selective Kondo scattering} The calculated spectral functions and orbital-projected density of states (DOS) of UTe$_2$ at $T$=25 K and 900 K are compared
in Fig~\ref{Fig_bands}a.
At 900 K, the U-5$f$-driven bands around -0.2 eV exhibit an incoherent feature, which shifts towards the Fermi level at 25 K, forming a strong Kondo peak with a kink band structure. 
This is reminiscent of the formation of a quasiparticle peak upon cooling, which leads to the formation of a Kondo cloud~\cite{kang_ndnio2,kang_topology}.
A distinctive feature of UTe$_2$ is that the U-5$f$-orbitals form a metallic band at high temperatures, reaching up to 6000 K~\cite{byung_ute2}.
UTe$_2$ is known as a high temperature Kondo lattice with a Kondo scale of approximately 500 K, as estimated from the behavior of the U-5$f$ DOS and local angular momentum susceptibility~\cite{byung_ute2}.

As shown in Fig~\ref{Fig_bands}a, the U-5$f$ peak is slightly below the Fermi level at 900 K, and the dispersive U-5$f$ metallic bands are extended to the Fermi level. This gives rise incoherent Kondo hybridization between U-5$f$ and conduction electrons of Te2-5$p$ and U-6$d$.
As shown in Fig.~\ref{Fig_bands}b, the incoherent Kondo hybridization between the Te2-5$p$ state of $|j= 3/2,j_{z}= -3/2\rangle$ and the dispersive U-5$f$ bands results in the appearance of dispersive Te2-5$p$ state of $|j= 3/2,j_{z}= -3/2\rangle$ bands along the $\Gamma$-Z direction at a temperature of 2000 K.
The dispersive band of Te2-5$p$ state $|j=3/2, j_z=-3/2\rangle$ also appears along the $\Gamma$-X and $\Gamma$-Y high symmetry lines.
These results indicate that incoherent Kondo scattering occurs in all directions from $\Gamma$, resulting in the formation of the 3D Fermi surface centered at $\Gamma$ at high temperatures.
Upon cooling, the coherent features of Kondo scattering become more pronounced, resulting in a brighter spectral weight. In particular, the coherent character at 25 K is strongly enhanced along the $\Gamma$-Z direction, which leads to the quasi-2D Fermi surface. This is observed as a cylindrical Fermi surface, as shown in Fig.~\ref{Fig_2DFS}d.

As shown in Fig.~\ref{Fig_bands}c, the incoherent Kondo hybridization between the U-6$d$ state of $|j= 3/2,j_{z}= -3/2\rangle$ and the dispersive U-5$f$ bands results in the appearance of dispersive U-6$d$ state of $|j= 3/2,j_{z}= -3/2\rangle$ bands along the $\Gamma$-X,Y, and Z  directions at a temperature of 2000 K.
Upon cooling, the coherent features of Kondo scattering become more pronounced, resulting in a brighter spectral weight in the vicinity of the Fermi level. In particular, the coherent character at 25 K is strongly enhanced along the $\Gamma$-Y direction, leading to the dumbbell shaped 3D Fermi surface, as shown in Fig.~\ref{Fig_3DFS}b.

\section*{Discussion}
We discovered that Kondo hybridization leads to Fermi surface reconstruction in UTe$_{2}$.
Interestingly, it is found that as the temperature decreases down to 25 K, the enhanced orbital selective Kondo coherence leads to the coexistence of a $\Gamma$-centered strong 3D Fermi surface and a quasi-2D Fermi surface, which gives a clear explanation for the observations of coexistence of 3D or 2D feature~\cite{christopher_prl2023}.

The two $\Gamma$-centered Fermi surfaces are comprised of conducting electrons from U-6$d$ and Te-5$p$. They are subject to orbital-selective Kondo scattering with U-5$f$. Our previous study in the same manner as this work, which predicted the Kondo scale using the calculated density of states and susceptibility, explains the anisotropic electronic resistivity measurements~\cite{byung_ute2, eo_prb2022}. This should be the evidence for the Fermi surface reconstruction originating from the Kondo effect. However, in the Kondo effect, since the localized U-5$f$ electrons dominate the Fermi surface, forming a peak in density of states at Fermi level at low temperatures, the spectral weight of both $\Gamma$-centered Fermi surfaces should be relatively weak, and their contribution to the Hall effect for the Fermi surface reconstruction should be limited~\cite{niu_prr2020}. Quantitative research on the effect of the two Fermi surfaces on the Hall effect demands further study.

\begin{figure}[ht]
\centering
\includegraphics[width=0.4
\textwidth]{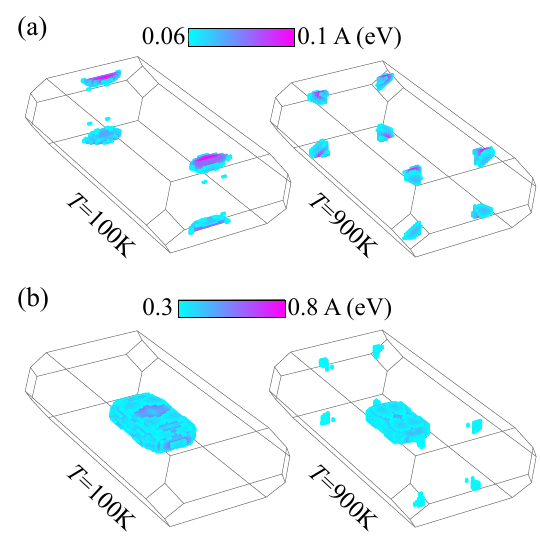}
\caption{\label{Fig_nso} The calculated orbital projected Fermi surface in the Brillouin Zone without spin-orbit coupling at 
$T$=100 and 900 K. (a) Te2-5$p$ state of $|l= 1,m= 0\rangle$ projected Fermi surface. (b) U-6$d$ states of $|l= 2,m= 2\rangle$ projected Fermi surface.
}
\end{figure}

Two DFT+DMFT studies on UTe$_{2}$ have been conducted, both of which do not take spin-orbit coupling (SOC) into account~\cite{xu2019quasi, Miao2020}. These studies reveal a flat U-5$f$ peak that is hybridized with conduction electrons at 10 K, while noting that the peak diminishes at 200 K.
This indicates that the Kondo scale in the non-SOC system is lower than 500 K in the SOC system~\cite{byung_ute2}.
By employing LQSGW+DMFT, our study revealed that in the absence of SOC, the uranium atom undergoes strong oxidation, with the U-5$f$ occupancy being 1.17, considerably lower than the 2.27 observed in the SOC simulation~\cite{byung_ute2}. This pronounced local magnetic moment is critical for Kondo scattering, implying that Kondo scaling is significantly reduced in the non-SOC system.
Figure~\ref{Fig_nso} show the Fermi surface of UTe$_{2}$ without considering SOC within LQSGW+DMFT.
In the non-SOC system, the U-6$d$ projected Fermi surface exhibits a 3D character. In contrast, the Te2-5$p$ projected Fermi surface shows no evidence of a $\Gamma$-centered Fermi surface. This suggests that SOC plays a crucial role in the reconstruction of the quasi 2D Fermi surface of Te2-5$p$ at low temperatures, as shown in Fig.~\ref{Fig_2DFS}, where the SOC is explicitly included.
The significant local U-5$f$ magnetic moment, originating from the strong SOC, leads to a flat U-5$f$ band at the Fermi level. At sufficiently low temperatures, the flat U-5$f$ band undergoes Kondo hybridization with Te2-5$p$ along the c-axis, leading to the quasi-2D Fermi surface. 

Although the spectral weight of both $\Gamma$-centered Fermi surfaces are weaker than that of the non-$\Gamma$-centered cylindrical Fermi surface, they should be observable for a short period of time. As the temperature decreases below 25 K, the Kondo coherence should become stronger. 
Consequently, it is expected that both the $\Gamma$-centered 3D and quasi-2D Fermi surfaces should be more observable and have longer lifetimes below 25 K, which our methods can't access due to the sign problem in Monte Carlo simulations.

Our results show that the coexistence behavior of 3D and quasi-2D Fermi surfaces can be clearly understood by the evolution of orbital selective Kondo hybridization at low temperature. It gives an explanation to inconsistent experimental observations due to orbital and temperature dependent Fermi surfaces.
The $\Gamma$-centered 3D Fermi surface should potentially host topological superconductivity, as has been suggested. The impact of the quasi-2D Fermi surface centered around $\Gamma$ on topological superconductivity should be investigated in future studies. 



\section*{Methods}
\subsection{LQSGW and DMFT calculations\\}

The electronic structure of UTe$_{2}$ was calculated using the $ab$-$initio$ linearized quasiparticle self-consistent GW (LQSGW) combined with dynamical mean field theory (DMFT)~\cite{tomczak2015qsgw,choi2019comdmft}. Classified under the Immm (No. 71) space group, this material's structure has been confirmed by previous investigations~\cite{stowe1996contributions,ikeda2006single}. The The LQSGW+DMFT approach is a streamlined variant of the GW+EDMFT method~\cite{sun2002extended,biermann2003first,nilsson2017multitier,kangfgwedmft,kang_fese}, wherein LQSGW techniques~\cite{kutepov2012electronic,kutepov2017linearized} are employed to derive the electronic structure. 
DMFT is then used to correct any discrepancies in the local GW self-energy~\cite{georges1996dynamical,metzner1989correlated,georges1992hubbard}.
In this study, we utilized experimentally measured lattice constants ($a=$ 4.1611, $b=$ 6.1222, $c=$ 13.955 $\textrm{\AA}$)~\cite{ikeda2006single} and computed all other necessary parameters, including double-counting energy and the Coulomb interaction tensor. Local self-energies for U-5$f$ and U-6$d$ were derived by solving two distinct single impurity models using the continuous time quantum Monte Carlo (CTQMC) method. The LQSGW+DMFT framework was implemented with the ComDMFT code~\cite{choi2019comdmft}, while the LQSGW calculations were performed using the FlapwMBPT code~\cite{kutepov2017linearized}. Further details can be found in the Supplementary Methods section.

\bigskip

\section*{Acknowledgments} 

We acknowledge the High Performance Computing Center (HPCC) at Texas Tech University for providing computational resources that have contributed to the research results reported within this paper.
C.H. Park acknowledges the support by the National Research Foundation of Korea (NRF) grant (Grant No. NRF-2022R1A2C1005548).

\bigskip
\textbf{Competing Interests} The authors declare no competing interests.

\bigskip
\textbf{Data availability} The data that support the ﬁndings of this study are available from the corresponding
authors upon reasonable request.


\bigskip
\textbf{Author contributions} B.K. designed the project. B.K. performed the calculations and conducted the data analysis. All authors wrote the manuscript, discussed the results, and commented on the paper.

\bibliography{ref}

\end{document}